\documentclass[a4paper,10pt]{article}
\usepackage{graphicx}

\newcommand{\alr}{\alpha_R}

\newcommand{\lnQ}{\ln(Q^2/\Lambda^2)}
\newcommand{\llnQ}{\ln\{\ln(Q^2/\Lambda^2)\}}
\newcommand{\llnt}{\ln\{\ln(t/\Lambda^2)\}}
\newcommand{\XL}{\Lambda^2}
\begin{document}
{\Large{\bf \noindent Analysis of pion electromagnetic
 form factor by the dispersion relation with QCD constraint}}
\vspace{32pt}

\noindent {\large Keiji Watanabe}$^1$\footnote{e-mail:
watanabk@phys.meisei-u.ac.jp}, 
{\large Hirohisa Ishikawa}$^2$\footnote{e-mail: ishikawa@meikai.ac.jp},
 {\large Masami Nakagawa}$^3$\\
\vspace{12pt}

\noindent ${}^1$ Department of Physics, Meisei university, Hino Tokoy
191-8506, Japan\\
${}^2$ Department of Economy, Meikai University, Urayasu Chiba,
279-8550, Japan\\
${}^3$ Deceased

\begin{abstract}
The pion electromagnetic form factor is investigated by using the
dispersion
 relation with the superconvergence condition, which was proposed to 
 synthesize the vector meson dominance model and QCD. The absorptive 
 part is given as an addition of the resonance and the QCD 
 terms, the latter being parametrized  
  so as to realize the prediction of perturbative QCD. 
  
  The experimental data are analyzed and the parameters for the QCD 
  part, the resonance part as well, are determined. Existing experiments 
  for the form factor are reproduced
 very well both for the space-like and time-like momenta.
 
\vspace{12pt}

\end{abstract}

\section{Introduction}
According to the perturbative QCD hadron electromagnetic form factors
decrease more
rapidly than an inverse polynomial for large squared momentum transfer
\cite{bf,h}. The boson electromagnetic form factor, $F(t)$, decreases as
\begin{equation} 
F(t)\to - 32\pi^2f_B^2/\beta_0t\ln|t| \label{qcd}
\end{equation}
 for $|t|\,\to \infty$,
where $f_B$ is the boson decay constant, $t$ is the squared momentum
 transfer and $\beta_0$ is the lowest order of the $\beta$ function in
the 
 renormalization group, namely $\beta_0\,= \,11-2n_f/3$ with $n_f$ being
the 
 number of flavor.
 Experimentally, the charge and magnetic form factors of nucleons 
 decrease more rapidly for
 large $t$ than the well-known dipole formula as is predicted by the 
 perturbative QCD. 
 
The QCD property for the form factor is realized by using the
dispersion relation in which the absorptive part, $\mbox{Im} F(t)$,
decreases 
 as some power of $\ln t$ and satisfies the superconvergence constraint 
\cite{dubnicka,fw}.  We break up the absorptive part as a 
 simple summation of the resonance and the QCD part. The former is 
 given as addition of the Breit-Wigner formula with possible mixing 
 among resonances and the latter as a power series in the running
coupling 
 constant which is analytically extrapolated to the time-like momentum. 
We impose the superconvergence condition on $\mbox{Im} F(t)$. 
 For the form factor the low momentum part is rather easy 
to deal with by the dispersion relation. 
It is, therefore, possible to treat whole range, from the low to 
the high momentum regions, systematically; the vector meson dominance
model (VMD)
 is realized for low momentum and the QCD condition is satisfied
  for large $|t|$. The dispersion relation works as an interpolation.
 
 In this paper we improve the result of Ref.\cite{nw1}, by investigating 
 correlations among QCD parameters. We search much wider range of
parameters
  than in Ref.\cite{nw1}. 
  We show that QCD part of $\mbox{Im}F(t)$ can be
 determined from the experiments on the pion electromagnetic form factor 
 although the data are restricted to low momentum. At least the second 
  order in the running coupling constant is necessary to fit the
experimental 
  data. The QCD scale parameter $\Lambda$ is determined to be in the
range 
 $\Lambda$ = $0.3\,\sim \,2$ GeV.  The 
 calculated form factor agrees with the existing experimental results
very well 
 both for the time-like and space-like momenta. 

\section{Superconvergence dispersion relation and QCD}
As the superconvergent dispersion relation has already been 
discussed \cite{dubnicka}-\cite{nw2}, 
 we  summarize our prescription to make the paper 
self-contained. To realize the asymptotic form of QCD we make use of the 
following property of the Hilbert 
transformation. Let $F(t)$ be given as
\begin{equation}
F(t)=\int_{s_0}^{\infty}dt'\frac{\rho(t')}{t'-t}, \label{hilbert}
\end{equation}
where $s_0\,>\,0$ is a constant and the function $\rho(t')$ decreases as
$$
\rho(t')\,\to \,a/\{\ln(t')\}^{\gamma}\quad
\,\,\mbox{for}\,\,t'\,\to\,\infty,
$$
with $a$ and $\gamma$ being constants, then 
$$
F(t)=\int_{s_0}^{\infty}dt'\frac{\rho(t')}{t-t'}\,\to 
\frac{a}{(\gamma-1)\{\ln(|t|)\}^{\gamma-1}} \quad 
\,\,\mbox{for}\,\,t\,\to\,\infty.
$$
Here $\gamma\,>\,1$. In the case that the spectral 
function $\rho(t')\,\to 
a/t'\ln^{\gamma}(t')$ for $t'\, \to\,\infty$ and satisfies the
superconvergence 
condition $\displaystyle{\int_{s_0}^{\infty}\rho(t') dt'\,=\,0}$, 
$F(t)$ given by (\ref{hilbert}) decreases asymptotically as   
 $F(t)\,\to \,a/(\gamma-1)t\,\ln^{\gamma-1}(|t|)$ for 
$t\,\to$ $\infty$. 
\subsection{Vector meson dominance model and QCD}
For the pion electromagnetic form
 factor, $F(t)$, we assume the unsubtracted dispersion relation with
respect to 
 the squared momentum transfer $t$. 
\begin{equation}
F(t)=\frac{1}{\pi}\int_{s_0}^{\infty}dt'\frac{\mbox{Im}F(t')}{t'-t},
\end{equation}
where $s_0$ is the threshold.
 To satisfy the asymptotic condition of perturbative QCD (\ref{qcd}), 
 the absorptive part should satisfy the following conditions:
\begin{equation}
\mbox{Im}F(t)\to -\frac{32\pi^3f_{\pi}^2}{t[\beta_0\ln^2(t/\Lambda^2)]}
\quad (t\to\infty)                                      \label{asympt}
\end{equation}
and the superconvergence condition
\begin{equation}
\int_{s_0}^{\infty}dt'\mbox{Im}F(t')=0. \label{super}
\end{equation}
 We take $\mbox{Im}F$ so as to realize the result of the vector meson
dominance
model for low momentum region and the QCD prediction asymptotically; we
express it as the summation of the Breit-Wigner formula  with mixing
among
 resonances \cite{bernicha} and the QCD term.
\begin{equation}
\mbox{Im}F(t)=\mbox{Im}F^{BW}(t)+\mbox{Im}F^{QCD}(t),
\end{equation}
where the Breit-Wigner term $\mbox{Im}F^{BW}(t)$ is given as
\begin{eqnarray}
\mbox{Im}F^{BW}(t)&=&\sum_j\frac{c_jM_j^2\gamma_j}{(M_j^2-t)^2+\gamma_j^2}
\nonumber\\
&&\hspace{-12pt}+\sum_{j<k}c_{jk}\Big[
\frac{\alpha_{jk}^I(M_j^2-t)+\alpha_{jk}^R\gamma_j}{(M_j^2-t)^2+\gamma_j^2}
-\frac{\alpha_{jk}^I(M_k^2-t)+\alpha_{jk}^R\gamma_k}{(M_k^2-t)^2+\gamma_k^2}
\Big].                                                         
\label{bw}
\end{eqnarray}
Here the suffixes denote numbering of resonances and $\alpha_{jk}^R$ and
 $\alpha_{jk}^I$ are
\begin{eqnarray}
\alpha_{jk}^R&=&-\frac{\gamma_j\gamma_k(M_j^2-M_k^2)}{(M_j^2-M_k^2)^2+
(\gamma_j-\gamma_k)^2},\\
\alpha_{jk}^I&=&-
\frac{\gamma_j\gamma_k(\gamma_j^2-\gamma_k^2)}{(M_j^2-M_k^2)^2+(\gamma_j-\gamma_k)^2}.
\end{eqnarray}
$\mbox{Im}F^{QCD}(t)$ is given as a power series in the running coupling
constant
multiplied by a function $h(t)$, which is introduced to assure the
threshold
 behavior and the convergence of the superconvergence integral 
 (\ref{super}). In fact,
\begin{equation}
\mbox{Im}F^{QCD}(t)=\mbox{Im}\Big[\sum_{j\ge1}
c_j^{QCD}\{\alpha_R(t)\}^j\Big]h(t).           \label{ImFQCD}
\end{equation}
Here $\alpha_R$ is the running coupling
 constant in QCD analytically regularized \cite{nw1,dw,dmw}. We shall 
 discuss on $\alpha_R$ in the next subsection.
   For the function $h(t)$ we assume the following formula \cite{nw2},
  motivated by Kroll et al.\cite{kroll}:
\begin{equation}
 h(t)=\left[\frac{t-t_0}{t+t_1}\right]^{3/2}\frac{t_2}{t+t_2},
\end{equation}
 with $t_i$ $(i=0,1,2)$ being parameters. 
It must be noticed that the result is not sensitive to the form of 
the function $h(t)$.

For the form factor we impose the normalization condition at 
$t\,=\,0$,
\begin{equation}
     F(0)=1.           \label{norm}
\end{equation}
We have two conditions on $F(t)$, the superconvergence and the 
normalization condition (\ref{super}) and (\ref{norm}), respectively.

\subsection{Running coupling constant for the time-like momentum}
It is known that the running coupling constant $\alpha_S$ has an
unphysical
pole for the space-like momentum if it is calculated by the perturbative
QCD. The removal of the 
singularity was proposed by Dokshitzer et al.\cite{dw,dmw} via 
analytic regularization.  The regularized running coupling constant,  
being denoted as $\alpha_R$, is given by the spectral representation
\begin{equation}  
\alpha_R=\frac{1}{\pi}\int_0^{\infty}dt'\frac{\sigma(t')}{t'-t}.\label{alR}
\end{equation}
Here the function $\sigma$ is given in terms of the $\alpha_S$, 
which is obtained for the space-like momentum via the renormalization
group. 
We have
\begin{equation}
\sigma(t)=\frac{1}{2i}[\alpha_S(e^{-i\pi}t)-\alpha_S(e^{i\pi}t)]. 
                                                      \label{sigma}
\end{equation}
We use the three loop approximation for $\alpha_S$, which is 
expressed by the formula
\begin{eqnarray}
\alpha_S(Q^2)&=&\frac{4\pi}{\beta_0}\Big[\lnQ+a_1\llnQ \nonumber \\
&&+a_2\frac{\llnQ}{\lnQ}+\frac{a_3}{\llnQ}+\cdots\,\Big]^{-1} 
                                                        \label{alS}
\end{eqnarray}
where
\begin{equation}
a_1=2\beta_1/\beta_0^2,\, a_2=4\beta_1^2/\beta_0^4,\, 
a_3=(4\beta_1^2/\beta_0^4)(1-\beta_0\beta_2/8\beta_1^2), \label{ai}
\end{equation}
 and  
$\beta_i\,(i=0,1,2)$ are the $\beta$ functions in QCD with 
$\beta_0=11-2n_f/3,\, \beta_1=51-19n_f/3$ and 
 $\beta_2=2857-5033n_f/9+325n_f^2/27$, $n_f$ being the number of flavor. 
 Expanding (\ref{alS}) in terms of $\lnQ$ and $\llnQ$, we get 
 the usual expression for the running coupling constant \cite{rev}. 
We perform the analytic continuation  of the  effective coupling
constant 
$\alpha_S$ to the time-like momentum by the replacement 
$Q^2\to e^{-i\pi}t$ and decompose to the real and imaginary parts
$\alpha_S=\mbox{Re}[\alpha_S(t)]$ + $i\, \mbox{Im}[\alpha_S(t)]$.
Substituting 
$\mbox{Im}[\alpha_S(t)]$
 in (\ref{sigma}), we obtain the running coupling constant for the 
 time-like momentum. We write it by the same notation as $\alpha_R$.  
  We showed in Ref.\cite{nw2} that $\alpha_R$ 
 given by (\ref{alR}) approximately coincides with that which is given
by
  (\ref{alS}) 
 with the ghost pole subtracted, namely, for the space-like momentum
\begin{equation}
\alpha_R(Q^2) \approx \alpha_S(Q^2)-A^{*}/(Q^2-Q^{*2}), \label{alR1}
\end{equation}
where $Q^{*2}$ is the pole of $\alpha_S(Q^2)$ and $A^{*}$ is the 
residue  at the pole. Writing $Q^{*2}/\Lambda^2$ = $e^{u^{*}}$, we have
$u^{*}$ = 0.7910487 for $n_f$ = 2 with the three loop approximation and
$$
A^{*}=4\pi\Lambda^2e^{u^{*}}/\Big\{\beta_0\Big(1+\frac{a_1}{u^{*}}-
a_2\frac{\ln u^{*}}{u^{*\,2} }+\frac{a_2-a_3}{u^{*\,2}}\Big)\Big\},
$$
where $a_i$ are defined by (\ref{ai}). 
To extend to time-like momentum we replace 
$Q^2\,\to e^{-i\pi}t$ as is mentioned above. The difference between 
$\alpha_R$ given by the spectral representation and the approximate one 
is less than 
0.4 {\%} for $|t|$ $\stackrel{{}_>}{_{\sim}}$ 3 GeV$^2$. The 
approximation becomes better as $|t|$ becomes larger \cite{nw2}. 
 We shall use the approximate one for the regularized running coupling 
 constant hereafter; as we have 
multiplied the function $h(t)$ to define $\mbox{Im}F^{QCD}$, the 
contribution from the low momentum part is considerably suppressed 
so that the approximation turns out to be very good. 

For the time-like momentum $\alr$, being obtained by the replacement 
$Q^2\to e^{-i\pi}t$ in (\ref{alR1}), is given as follows:
\begin{eqnarray}
Re[\alpha_R(t)] &=&\frac{4\pi u}{\beta_0 D(t)}+\frac{A^{*}}{t+Q^{*2}}, 
                                                            \label{rear}
\\
Im[\alpha_R(t)] &=& \frac{4\pi v}{\beta_0 D(t)},              
                                                           \label{imar}         \end{eqnarray}
where
\begin{eqnarray}
u &=& \ln(t/\XL)+\frac{a_1}{2} \llnt 
\nonumber \\
&&+\frac{a_2}{\ln^2(t/\XL)+\pi^2}
\Big\{\frac{1}{2}\ln(t/\XL){\llnt}+\pi\theta\Big\} \nonumber \\
&&+\frac{a_3\ln(t/\XL)}{\ln^2(t/\XL)+\pi^2} \\
v &=& \pi+a_1 \theta -\frac{a_2}{\ln^2(t/\XL)+\pi^2}
\Big\{\frac{\pi}{2}\llnt-\theta \ln(t/\XL) \Big\}\label{v} \nonumber 
\\
&&-\frac{\pi a_3}{\ln^2(t/\XL)+\pi^2},
\end{eqnarray}
and
\begin{equation}
D=u^2+v^2.                           \label{D}
\end{equation}
Here 
\begin{equation}
\theta = \tan^{-1}\Big\{\frac{\pi}{\ln(t/\XL)}\Big\}. 
\end{equation}
The spectral function $\sigma$ is obtained by taking imaginary part of 
$\alpha_S$, that is, 
\begin{equation}
\sigma(t)=4\pi v/\beta_0 D,                         \label{s(t)}
\end{equation} 
 where 
 $v$ and $D$ are defined by (\ref{v}) and (\ref{D}), respectively.

\section{Numerical results}
We analyzed the experimental data of the pion electromagnetic form 
factor for the space-like and the time-like momenta. The parameters 
appearing in our formulas are determined so as to minimize the $\chi^2$, 
 addition of the chi square for the time-like and 
space-like momenta. We fix the parameters which appear in the function 
$h(t)$ at $t_0$ = $\Lambda^2$ and $t_1$ = $t_2$ =16 GeV$^2$ as the 
result is not sensitive to these values \cite{nw2}.
The number of flavor is taken as $n_f$ = 2.

For large $t$, $\mbox{Im}F$ given by (\ref{ImFQCD}), approaches to 
$\mbox{Im}F(t)\,\to\,c_1^{QCD}\mbox{Im}\alr(t)t_2/t$ and 
$\mbox{Im}\alr$ given by (\ref{imar}) to $4\pi^2/\beta_0 
\ln^2(t/\Lambda)$. 
Therefore, comparing (\ref{asympt}) and (\ref{ImFQCD}), we have
\begin{equation}
c_1^{QCD}\,=\,-8\pi f_{\pi}^2/t_2,       \label{cQCD1}
\end{equation}        
which is fixed at
 $c_1^{QCD}$ = $-0.02683$ by taking $t_2$ = 16 GeV and
  the pion decay constant $f_{\pi}$ = 0.1307 GeV.

To realize the data for the time-like momentum we take the masses and 
widths of vector bosons as parameters, which are determined as follows:
$m_{\rho}$ = 0.760 GeV, 
 $\Gamma_{\rho}$ = 0.138 GeV; $m_{\rho'}$ = 1.40
GeV, $\Gamma_{\rho'}$ = 0.33 GeV; $m_{\rho''}$ = 1.73 GeV, 
$\Gamma_{\rho''}$ = 0.24 GeV. We investigated the effect of the
unconfirmed vector
 boson whose mass and width are taken as 
  $m_{\rho'''}$ =2.11 GeV and $\Gamma_{\rho'''}$ = 0.368 GeV.
  The mass and width of $\omega$ meson are kept at the experimental
values
 \cite{rev}. We leave out the $\phi$ meson in this calculation as the 
 contribution is estimated to be very small. In the chi square 
 analysis the following 
 experimental data are used: Amendolia et al.\cite{am} and Bebek et 
 al.\cite{bebek} 
 for the space-like momentum and Barkov et al.\cite{barkov} for the 
 time-like momentum. The number of data points  are 101 for the
space-like and
  157 for the time-like regions;
 totally 258 data are used in the analysis. Degrees of freedom (DOF) 
 is 241 when $\rho'''$ is omitted. If $\rho'''$ is taken into account,  
 DOF = 238.

  First, we investigate the QCD scale parameter $\Lambda$. We restrict
to
  the second order approximation for the QCD running coupling 
  constant in $\mbox{Im}F^{QCD}$,
  namely, $c_3^{QCD}$ = 0.  We illustrate in Fig.1 the contour
   lines for $\chi^2$ = 400, 500 and 600 to 
  give correlation between the pion decay constant $f_{\pi}$, being 
  taken as a parameter, and $c_2^{QCD}$. Here  
  $\rho'''$ is left out. To draw the contour lines, $f_{\pi}$ and
   $c_2^{QCD}$ is kept fixed 
   and the other parameters are determined so as to make 
  $\chi^2$ minimum. The QCD scale parameter is fixed at $\Lambda$ = 
  0.8 GeV. $\chi^2$ is then given as a function of $f_{\pi}$ and 
  $c_2^{QCD}$. The horizontal lines in Fig.1 denote the experimental 
  value for the pion decay constant $f_{\pi}$ =$0.1307\pm 0.00037$ GeV. 
  In the same way, we examined correlation between $\Lambda$
  and $c_2^{QCD}$. In Fig.2 we show the contour lines which give 
  correlation between $\Lambda$, being taken as a parameter, and 
  $c^{QCD}_2$, where the pion decay constant is fixed at the
experimental 
  value, $f_{\pi}$ = 0.1307 GeV.
   Taking $\chi^{2}\,\leq 500$ in Fig.1 and the experimental value for 
   $f_{\pi}$, we obtain $c^{QCD}_2\,=\,-0.87 \sim -0.49$. 
   The one-loop approximation for the
   QCD part, namely $c^{QCD}_2$ = 0, leads to poor result.
  The QCD scale parameter is determined from the correlation of 
  $c_2^{QCD}$ and $\Lambda$ which is illustrated in Fig.2. The 
  condition  $\chi^{2}\,\leq\,500$ leads to  
  $0.3\,\stackrel{_{<}}{_{\sim}} \Lambda \,\stackrel{_{<}}{_{\sim}}\,2$ 
  GeV, where we take the value of $c^{QCD}_2$ determined above, 
  namely, $c^{QCD}_2\,=\,-0.87 \sim -0.49$.
  The best fit is obtained for $\Lambda\,\approx\,0.8$ GeV.  
 \vspace{24pt}
 
\begin{center}
\includegraphics[width=.5\linewidth]{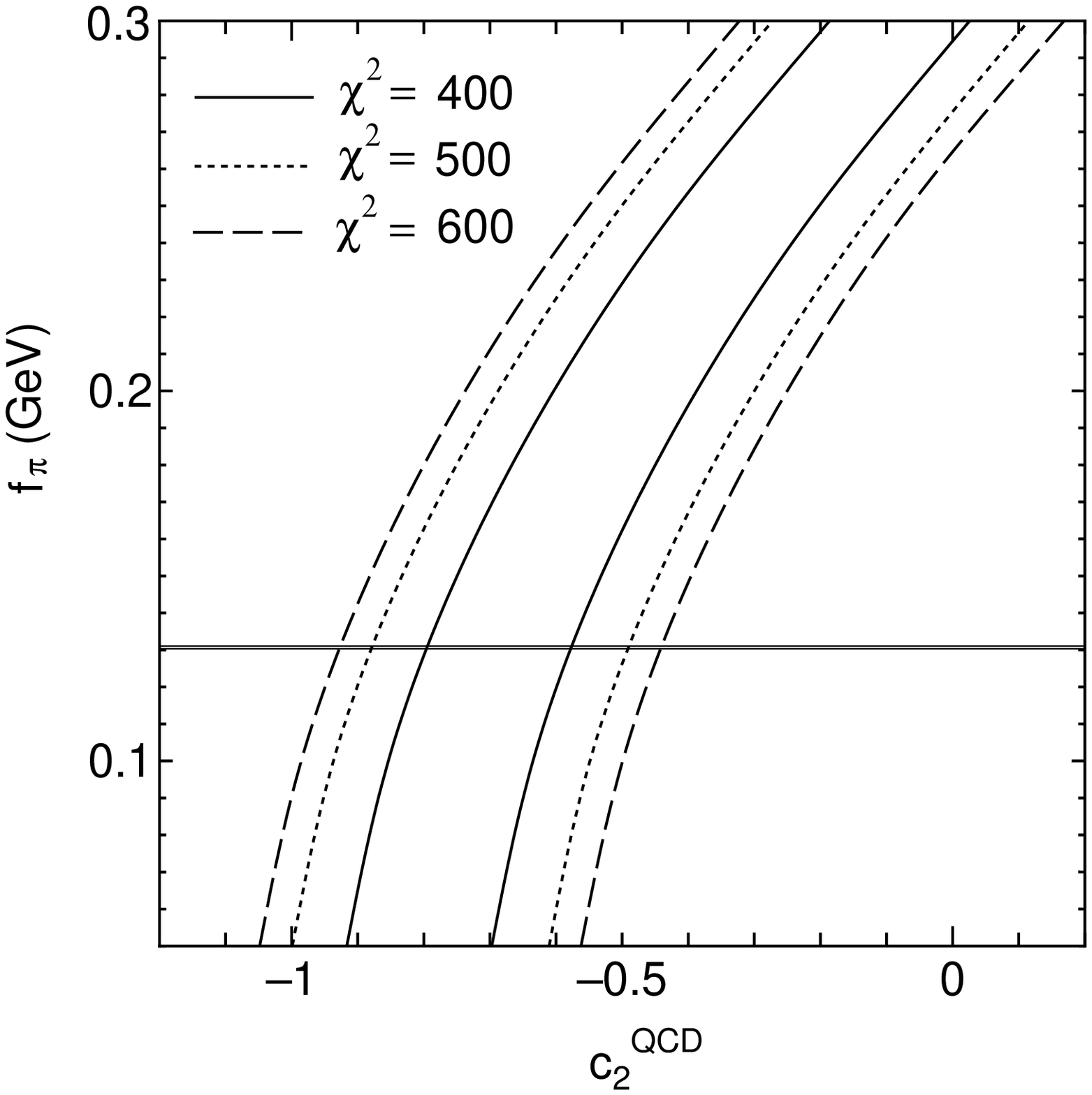}
\end{center}

{Fig.1. \noindent \small \noindent Contour lines for the parameters $c_2^{QCD}$ and $f_{\pi}$. 
$\chi^2$ is obtained 
as a function of $f_{\pi}$ and $c_2^{QCD}$ with the other parameters 
determined so as to make $\chi^2$ minimum. $\Lambda$ is fixed at 0.8 
GeV. Solid, dotted and dashed 
lines denote the contours with $\chi^2$ = 400, 500 and 600, 
respectively. Data for the space-like momentum \cite{am}, \cite 
{bebek} and the time-like momentum \cite {barkov} are used.}
\vspace{24pt}

\begin{center}
\includegraphics[width=.5\linewidth]{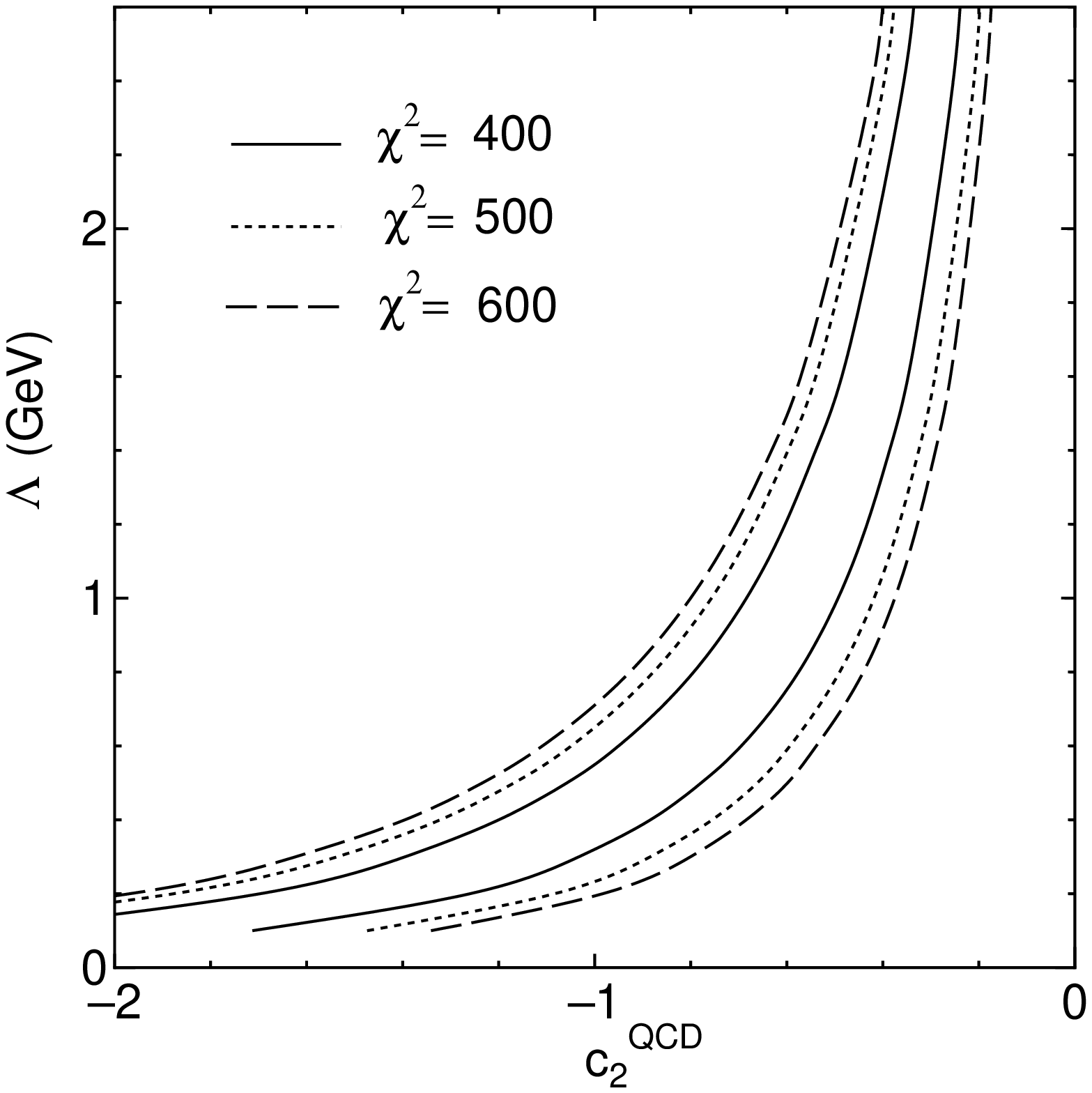}
\end{center}

{\small \noindent
Fig.2. Contour lines for $c_2^{QCD}$ and $\Lambda$. The pion decay constant 
is fixed at the experimental value. See the caption of Fig.1.}
\vspace{24pt}

The parameters which minimize $\chi^2$ are summarized in Table I, where
  $\Lambda$ is fixed at $\Lambda$ = 0.8 GeV and the pion decay 
  constant at the experimental value for the cases with and without 
  $\rho'''$.
  
\begin{center}
    \begin{tabular}{|c|c|c|}\hline
 {} & $\qquad$without $\rho'''(2.11)$ & with $\rho'''$(2.11)\\ \hline
$c_{\rho}$    &{\hspace{2pt}   2.723}       &{\hspace{2pt}    2.701
}               \\
$c_{\omega}$  &   $-$0.0030   &$-$0.0030                \\
$c_{\rho'}$   &  $-$0.2701    & $-$0.0924               \\
$c_{\rho''}$  &   $-$0.1956   & $-$0.2830               \\   
$c_{\rho'''}$ &   -----    &{\hspace{ 2pt}$-$0.0260}    \\
$c_{\omega'\mbox{-}\rho}$& 1.069     & 1.068            \\ 
$c_{\rho\mbox{-}\rho'}$  & 1.832     &{\hspace{-12pt} $-$0.089} \\
$c_{\rho\mbox{-}\rho''}$ &0.562      &{\hspace{-12pt} $-$0.136} \\
$c_{\rho'\mbox{-}\rho''}$&{\hspace{-26pt} $-$51.92}  &{\hspace{-24pt} 
$-$49.99} \\
$c_1^{QCD\,\dagger}$     &  $-$0.02683 & $-$0.02683               \\
$c_2^{QCD}$&  {\hspace{-12pt} $-$0.645 } & {\hspace{-12pt}$-$0.573 }\\
$\chi^2$   &  {\hspace{-24pt}379.1 }     & {\hspace{-24pt}355.1 }    \\ 
\hline
\end{tabular}
\end{center}
\hspace{3cm}$\dagger$ $c_1^{QCD}$ is calculated by (\ref{cQCD1}).

Table I {\small \noindent Residues at the resonance poles, mixing parameters among
 resonances and
the QCD parameters $c_i^{QCD}$ ($i =$ 1, 2). We take $\Lambda$ = 0.8 GeV
and 
$c_3^{QCD}$ =0. The parameters in the function $h(t)$ are $t_0$ =
$\Lambda^2$, 
$t_1$ = $t_2$ = 16 GeV$^2$. Two cases are investigated without the
unconfirmed 
resonance $\rho'''$ and with $\rho'''$. The value of $\chi^2$ is the 
summation of that for the space-like and time-like momenta.
 In Figs.1-5 the same parameters are used.}
\vspace{24pt}

Let us compare our calculated results with the experimental data of 
the pion electromagnetic form factor.
In Fig.3 we give our results for $|F(t)|^2$ for the space-like momentum. 
The experimental data are taken from Amendolia et al.\cite{am}. In 
Fig.4 $|t|\,F(t)$ is compared with the experiments. Our result agrees 
with the recent experiment by J.Volmer et al.\cite{volmer}. 
In the figure the data by 
Bebek et al.\cite{bebek} are included.

\begin{center}
\includegraphics[width=.5\linewidth]{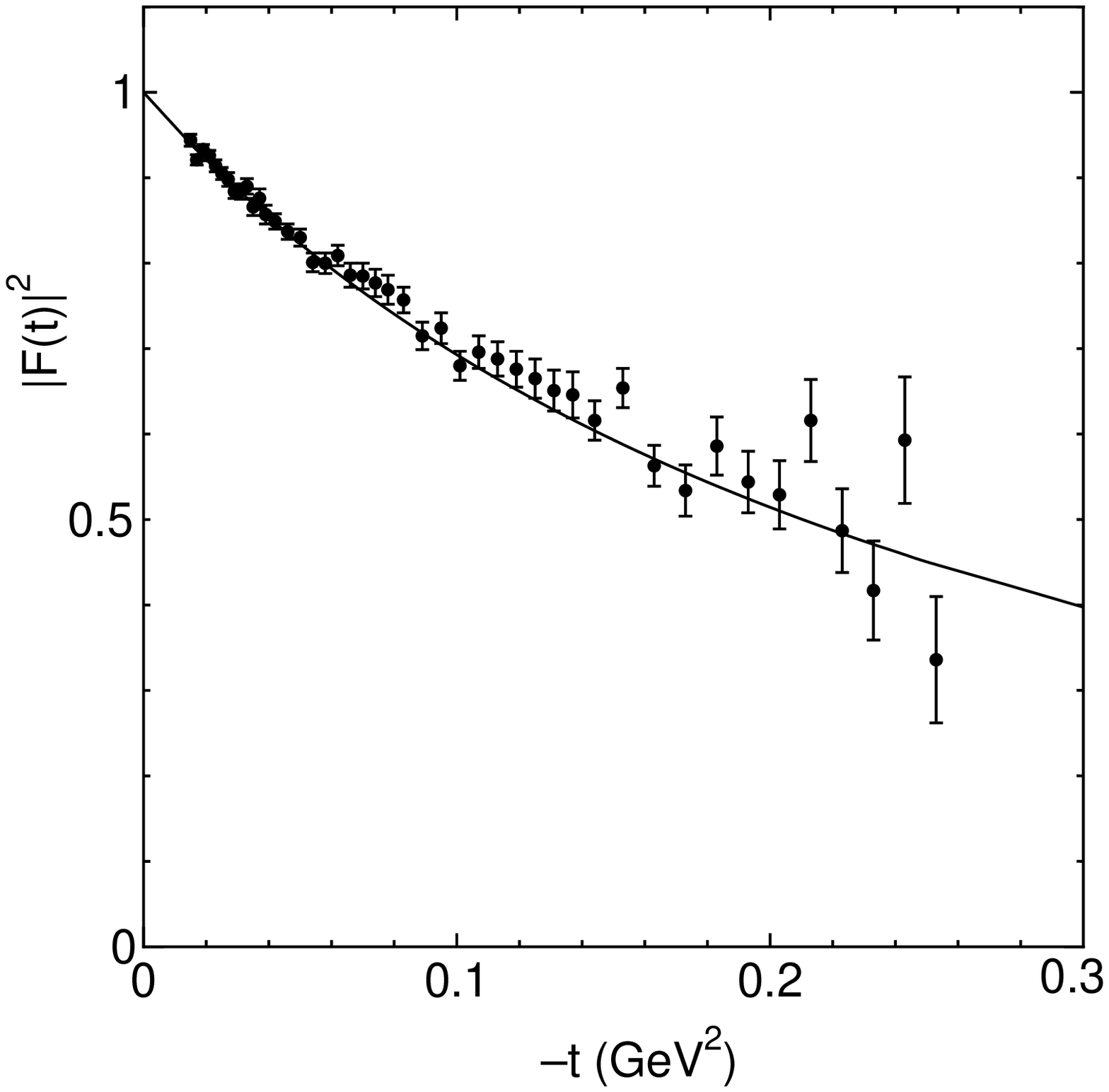}
\end{center}

{\small \noindent Fig.3. 
Electromagnetic form factor $|F(t)|^2$ for the space-like momentum. Low
momentum 
region. Data are taken from Amendolia et al.\cite{am}. Two cases with 
and without $\rho'''$ almost coincide. The parameters given in Table I
are used.}
\vspace{24pt}

\begin{center}
\includegraphics[width=.5\linewidth]{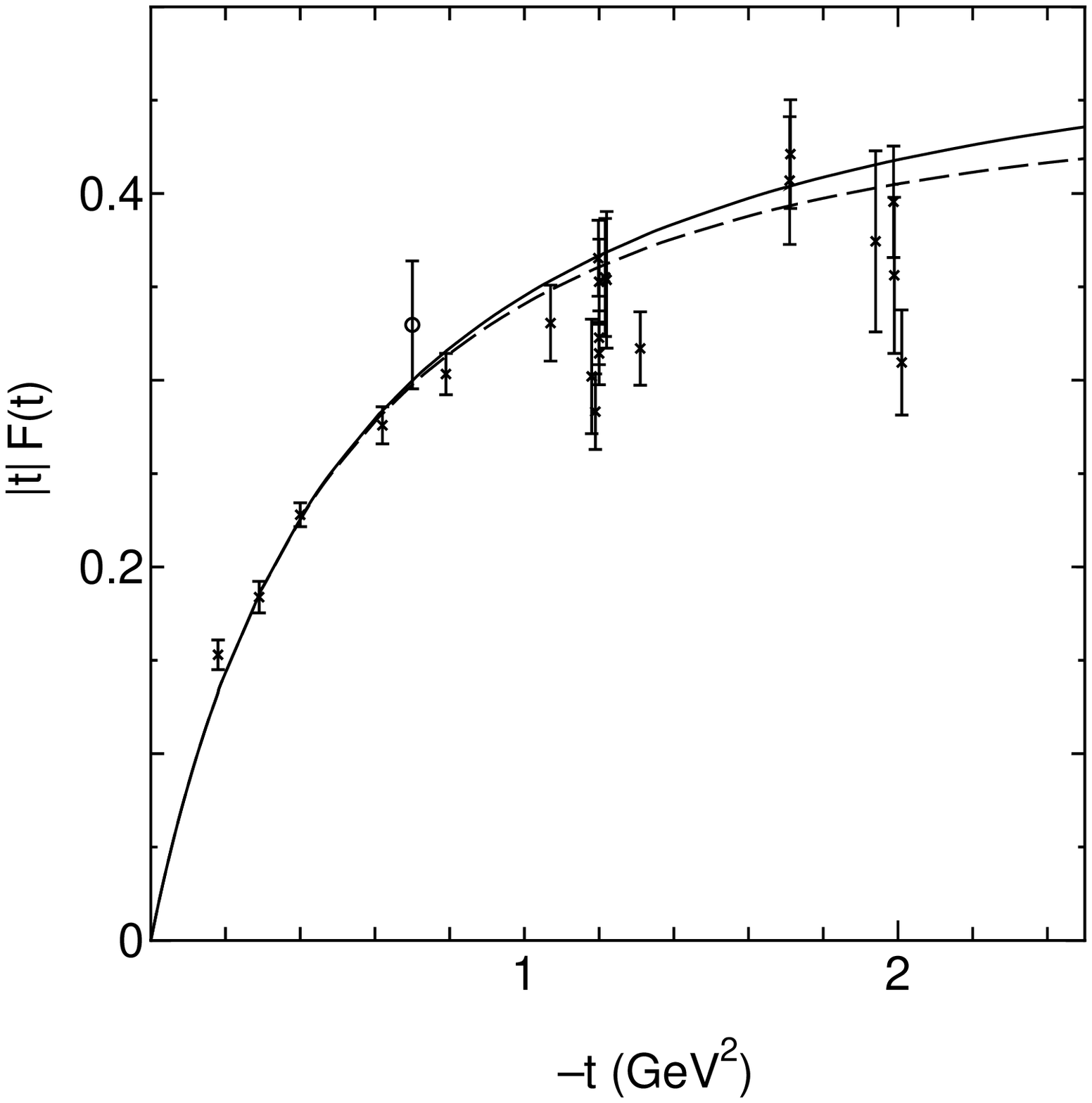}
\end{center}

{\small \noindent Fig.4. $t\,F(t)$ for the space-like momentum. The data denoted as 
$\times$ are 
taken from Bebek et al.\cite{bebek}, the open circles and closed circles
are 
taken from 
Volmer at al.\cite {volmer}. Solid line stands for the 
calculation without $\rho'''$ and the dashed one with $\rho'''$. The
parameters 
given in Table I are used.}
\vspace{24pt}

 The solid lines stand 
for the calculation without assuming 
$\rho'''$ and the dashed ones with $\rho'''$.

\begin{center}
\begin{tabular}{cc}
\includegraphics[width=.4\linewidth]{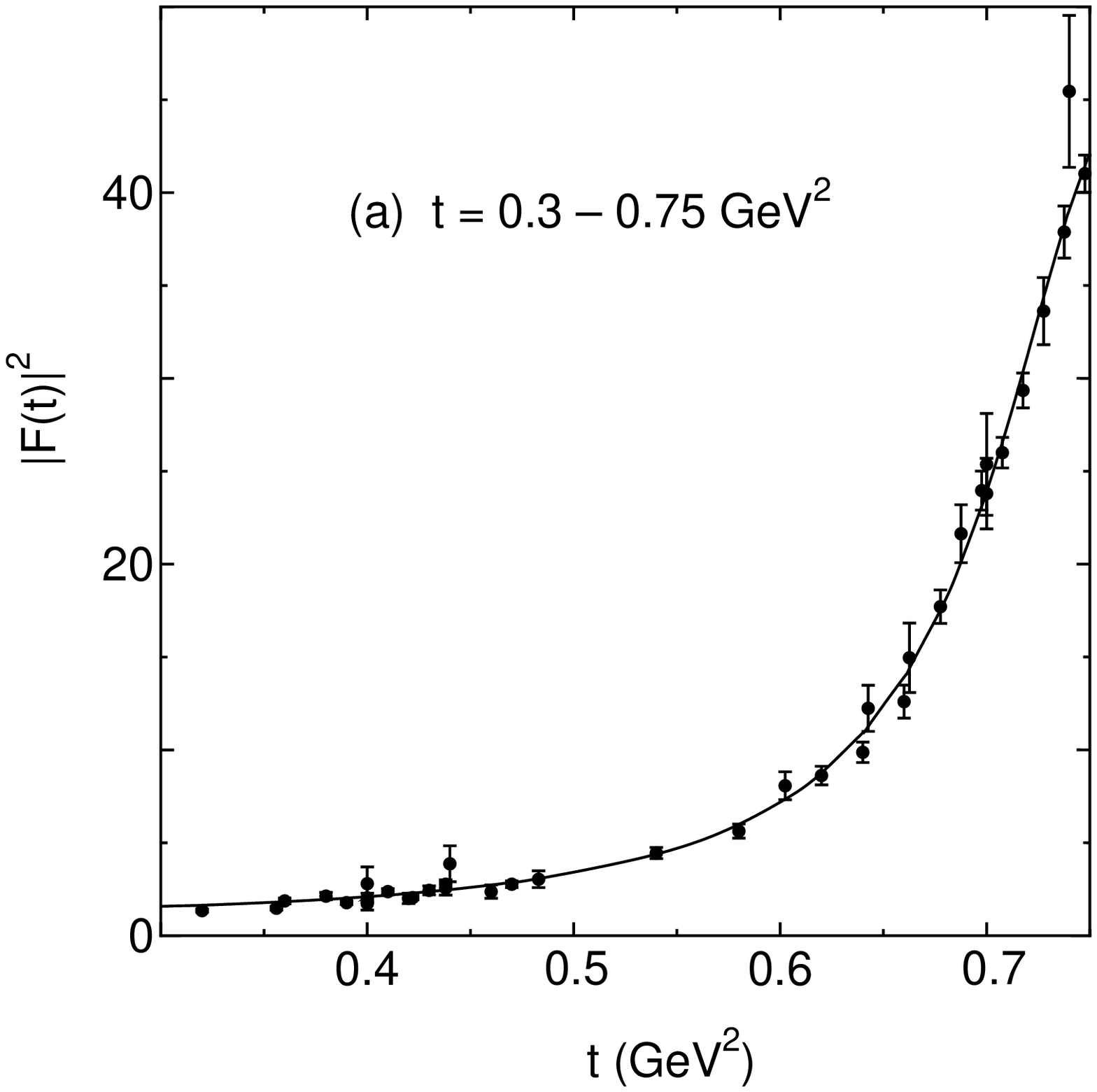}
&\includegraphics[width=.4\linewidth]{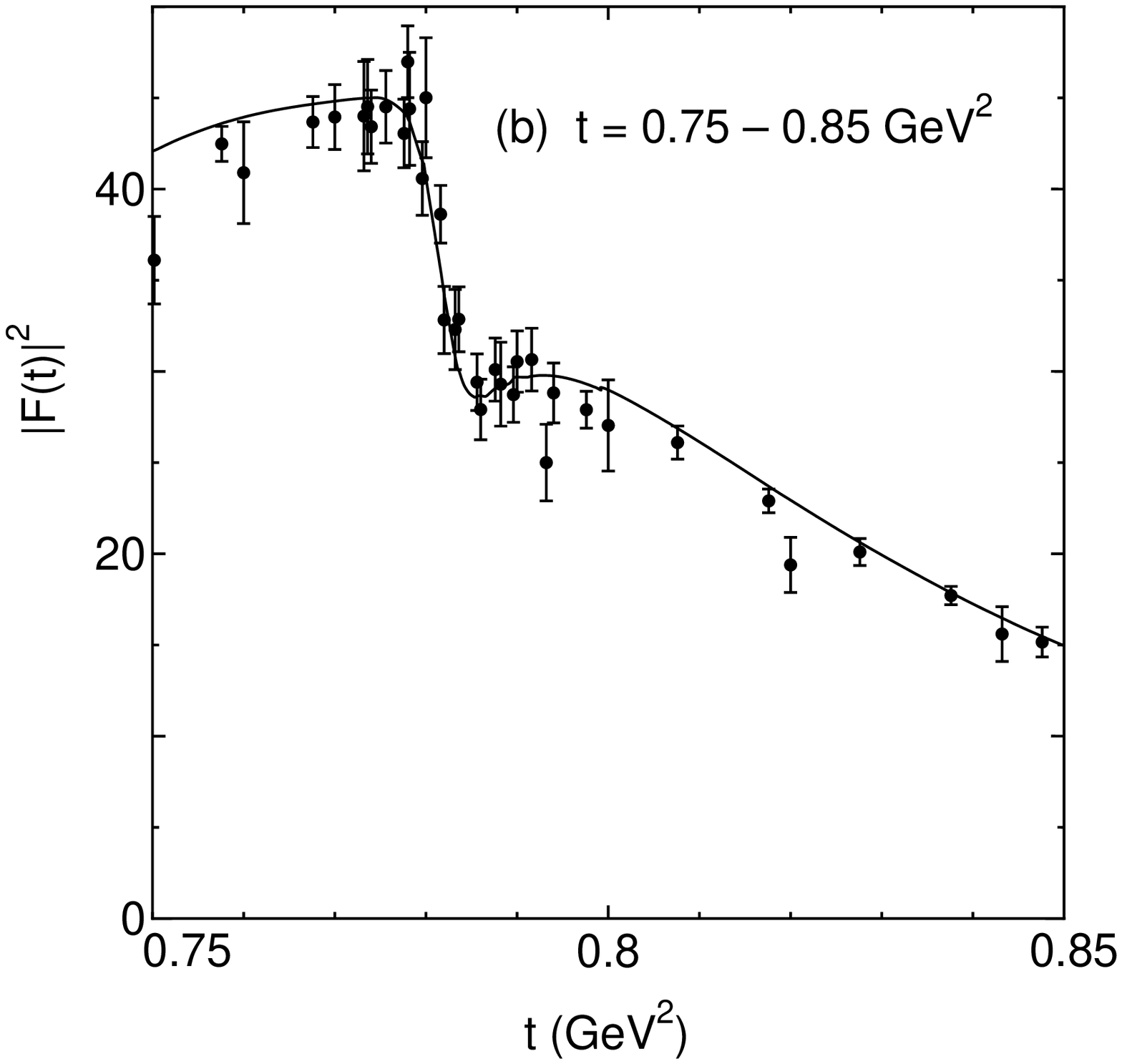}\\
\includegraphics[width=.4\linewidth]{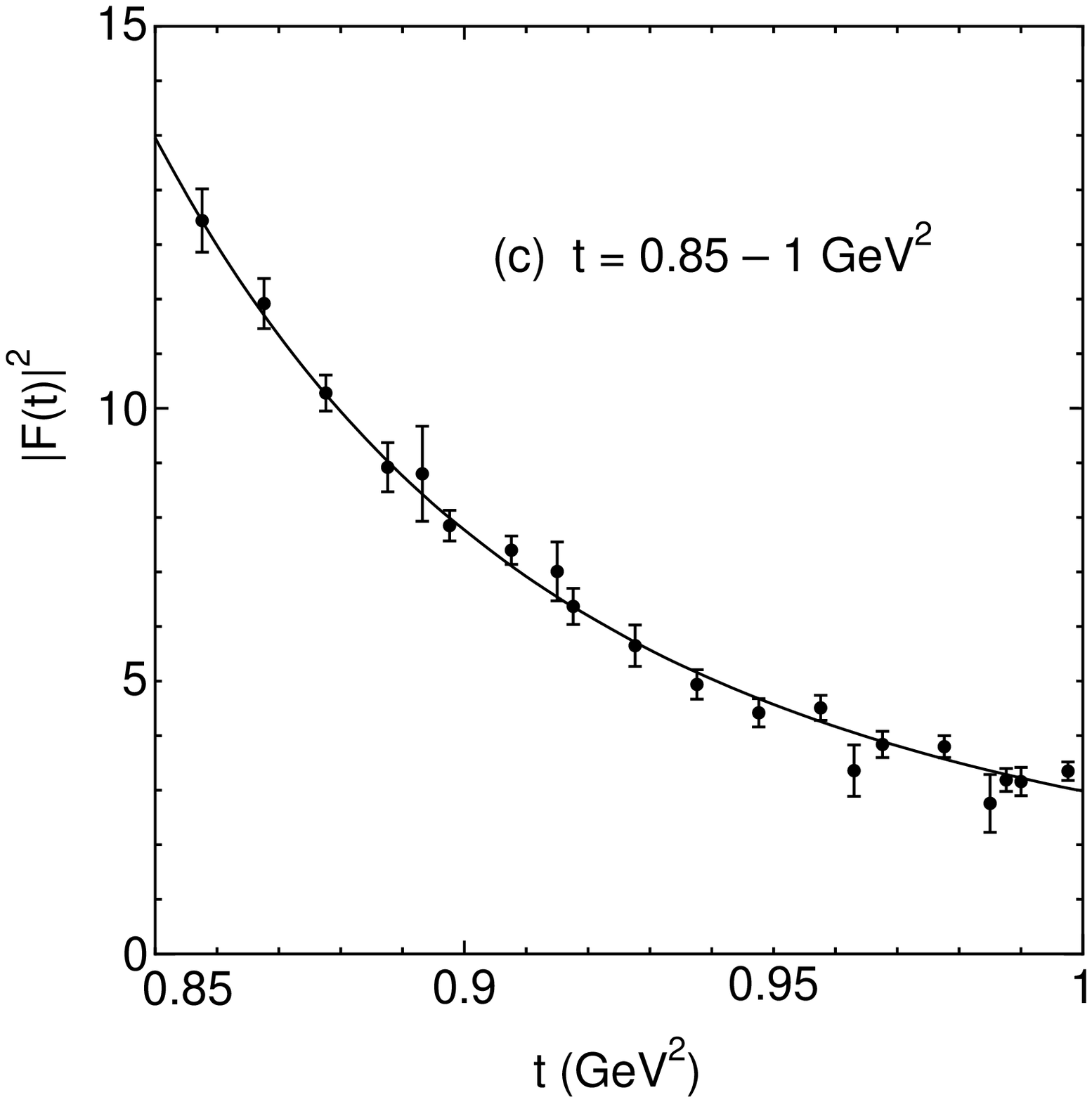}
&\includegraphics[width=.4\linewidth]{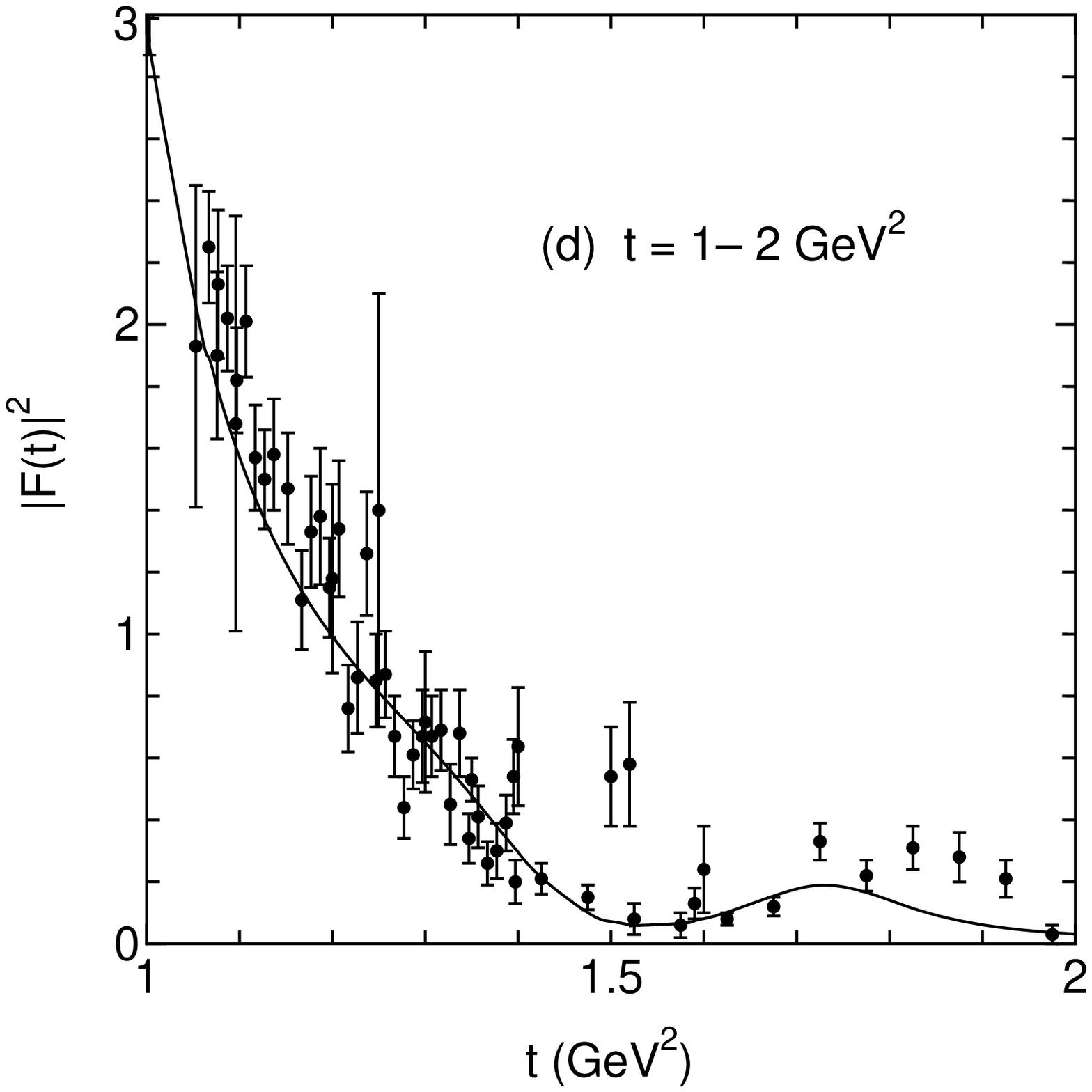}\\
\end{tabular}
\end{center}

{\small \noindent Fig.5 $|F(t)|^2$ for the time-like momentum. Data are taken from  
Barkov et al.\cite{barkov}. The solid lines stand for the calculation
without 
$\rho'''$ and the dashed ones with $\rho'''$. (a) $t$ = 0.3 $-$ 0.75 
GeV$^2$, (b) $t$ = 0.75 $-$ 0.85 GeV$^2$, (c) $t$ = 0.85 $-$ 1 GeV$^2$ 
and (d) $t$ = 1 $-$ 2 GeV$^2$. The parameters given in Table I are used.}
We compare in Fig.5 (a) - (d) the calculated results for $|F(t)|^2$ 
for the time-like momentum with the experimental data. It is necessary
to take 
account of the $\omega$ boson and $\rho$-$\omega$ mixing to explain the 
dip in the form factor near the 
$\rho$ meson mass. In Fig.4 and Fig.5 (a) - (d) the dashed curves denote the
result 
obtained by taking account of $\rho'''$. We find that the assumption of 
$\rho'''$ improves the result especially for the time-like momentum 
above 1.5 GeV$^2$. The chi square is obtained as $\chi^2$ = 379.1 for
the 
case without $\rho'''$ and $\chi^2$ = 355.1 for the case with $\rho'''$.

\section{Concluding remarks}
We analyzed the experimental data for the pion electromagnetic form 
factor by using the dispersion relation in which the QCD 
constraint is satisfied. The imaginary part of the form factor is given
as the 
summation of the Breit-Wigner formula with mixing among resonances and
the QCD term.
The latter is expressed as a power series in the running coupling
constant multiplied by 
 a function that assures the convergence of the 
superconvergence constraint. The function may be interpreted as
representing
 the quark-photon and  quark-hadron vertices. In order that the form
factor
  satisfies the QCD prediction we imposed the asymptotic condition on 
  the absorptive part, by which 
the coefficient of $\alpha_R$ in $\mbox{Im}F^{QCD}$, that is
$c_1^{QCD}$, is given 
in terms of the pion decay constant. The QCD constraint is 
very strong so that at least  $O(\alpha_R^2)$ approximation is required
to 
realize the experimental data. 
By taking the second order approximation for the QCD running coupling 
constant we are able to reproduce the experimental data of the pion
electromagnetic 
form factor very well. 
The result is improved by taking account of the third order term,
namely, 
 $c_2^{QCD}$ and $c_3^{QCD}$ as free parameters. However, the 
 experimental data are not sufficient to determine them accurately.  

We assumed the mixing among vector bosons. Especially mixing between 
$\omega$ and $\rho(760)$ is necessary to realize the structure of form 
factor near the $\rho$ meson mass as in Ref.\cite{bernicha}. The 
structure is explained by the mixing very well as is illustrated in
Fig.5 (b). 
An enhancement is observed 
in the pion electromagnetic form factor for $t$ = 1.1 $\sim$ 1.3 GeV
\cite{barkov} 
(see Fig.5(d)), although there is no 
isovector vector boson with the mass between 1.2 and 1.7 GeV. 
To explain the enhancement, without assuming 
resonance with the mass about 1.2 GeV, large mixing between $\rho'$ and
$\rho''$ is 
necessary. 

Analysis is performed by using the data of Amendolia et al.\cite{am} and
 Bebek et al.\cite{bebek} for the space-like 
momentum. As we have shown in Fig.3 our result agrees with the recent 
 experiments for the space-like momentum \cite{volmer}. 

We examined the effect of unconfirmed vector boson $\rho'''$ with the
mass 
above 2 GeV. It is shown that $\rho'''$ improves the result, especially 
for the time-like momentum with $t$ above 1.5 GeV$^2$ (see 
 Fig.5 (d)).

The QCD scale parameter $\Lambda$ is determined to be in the 
range $0.3\,\sim\,2$ GeV from the experiments of the pion 
electromagnetic form factor; the best fit is attained for$\ 
\Lambda\,\sim$ 0.8 GeV. $\Lambda$, obtained from the analysis of  
pion form factor, seems to be larger than that 
reported in the deep inelastic scattering. This implies that $\Lambda$
becomes 
larger when the number of effective flavor is decreased \cite{milton}.

\newpage
\noindent{\Large{\bf Appendix \hspace{6pt} A functional realization of 
theoretical results for the space-like momentum}}
\vspace{12pt}

For the convenience of reproducing the theoretical form factor for the
space-like 
momentum we 
express our calculated results by a functional approximation similar 
to the formula proposed by Gari and Kr\"umperman \cite{gk}, 
\begin{equation}
F(t)=\frac{M^2}{|t|+M^2}\,\frac{1+N_1 x+N_2 x^2}{1+D_1 x+D_2 
x^2+D_3x^3},
\end{equation}
where $x=\ln(1+|t|)$, with $t$ being expressed in the unit of GeV$^2$.
$M$, a constant with the dimension of mass, is fixed at the $\rho$ 
meson mass; $M$ = 0.76 GeV. The parameters $N_i$ and $D_i$ are 
determined so as to realize the calculated results given in Fig.4.

For the case without $\rho'''$, the parameters are determined as
follows:
$N_1 $ = 4.5919, $N_2$ = $-$0.62524, $D_1$ = 4.9604, $D_2$ = 
$-$0.7049, $D_3$ = 0.0540.
For the case with $\rho'''$ the parameters are 
$N_1$ = 4.7597, $N_2$ = 0.30520, $D_1$ = 5.1071, $D_2$ = 0.3598, $D_3$ = 
0.1904. The formula is valid up to $|t|\,\le\,$ 10 GeV$^2$; the
theoretical result is 
reproduced within 
the accuracy of 0.1{$\% $} for the squared momentum 
$|t|\,\le$ 10 GeV$^2$ and for $|t|\,\le 6$ GeV$^2$ the error is less 
than 0.02$\%$.

\end{document}